\documentclass[12pt]{article}
\usepackage{amssymb,amsmath}
\usepackage{latexsym}
\usepackage{graphicx}
\usepackage{psfrag}

\numberwithin{equation}{section}

\def\IR{\mathbb{R}}

\newcommand{\be}{\begin{equation}}
\newcommand{\ee}{\end{equation}}
\newcommand{\bea}{\begin{eqnarray}}
\newcommand{\eea}{\end{eqnarray}}

\def\nn{\nonumber \\}

\def\rm{\mathrm}
\newcommand{\Db}[1]{$\overline{\mathrm{D}#1}$}

\begin{document}

\begin{titlepage}

\begin{flushright}
DFPD11/TH-5 \\
AEI-2011-018
\end{flushright}

\bigskip
\bigskip
\centerline{\Large \bf A Black Ring with two Angular Momenta in Taub-NUT}
\bigskip
\bigskip
\centerline{{\bf Iosif Bena$^1$, Stefano Giusto$^{2,3}$ and Cl\'{e}ment Ruef$^{\, 4}$}}
\bigskip
\centerline{$^1$ Institut de Physique Th\'eorique, }
\centerline{CEA Saclay, 91191 Gif sur Yvette, France}
\bigskip
\centerline{$^2$ Dipartimento di Fisica ``Galileo Galilei,''}
\centerline{Universit\`a di Padova, Via Marzolo 8, 35131 Padova, Italy}
\bigskip
\centerline{$^3$ INFN, Sezione di Padova, Via Marzolo 8, 35131 Padova, Italy}
\bigskip
\centerline{$^4$ Max Planck Institute for Gravitation, Albert Einstein Institute}
\centerline{Am M\"uhlenberg 1, 14476 Golm, Germany}
\bigskip
\centerline{{ iosif.bena@cea.fr,~stefano.giusto@pd.infn.it,~clement.ruef@aei.mpg.de, } }
\bigskip
\bigskip

\begin{abstract}

We use the recently-constructed explicit duality transformation that relates a rotating \Db{6}-D4-D2-D0 black hole solution to a rotating M5-M2-P black string to construct a non-supersymmetric black ring in Taub-NUT that has two angular momenta, as well as M2 charges and M5 dipole moments. This is the first black ring solution that has both dipole charges and rotation along the $S^2$ of the horizon, and hence can be thought of as the ``Pomeransky-Senkov'' version of the M5-M2 black ring in Taub-NUT. Its physics should provide a testing ground for the applicability of the blackfold approach to charged rotating black branes, and should elucidate the phase space of charged dipole rings in various backgrounds. 

\end{abstract}

\end{titlepage}


\section{Introduction}

Besides providing explicit examples of black hole uniqueness violations in five dimensions, \cite{Emparan:2001wn,Emparan:2004wy,Elvang:2004rt,Bena:2004de,Elvang:2004ds,Gauntlett:2004qy} and revolutionizing our understanding of the phases of black hole solutions in various dimensions \cite{Elvang:2007hg,Emparan:2007wm}, black rings have triggered the crystallization of the so-called blackfold approach  \cite{Emparan:2009at,Emparan:2009zz} for determining whether a certain black membrane with certain angular momenta and a certain horizon topology exists in pure gravity in any number of dimensions. This approach is in the process of being extended to blackfolds with magnetic and electric charges in supergravity, which will help to better understand the physics of branes in thermal backgrounds and have important applications to the study of strongly-interacting gauge theories at finite temperature via the $AdS$-CFT correspondence \cite{Grignani:2011mr}.

One of the best benchmarks for testing and perfecting the blackfold approach and its charged generalizations is the construction of fully-backreacted explicit solutions that describe rings or membranes in various spacetimes. This is by no means an easy task, except when there exists an underlying principle (like supersymmetry) that allows one to simplify and solve the cohomogeneity-two Einstein's equations. Starting from an observation of Goldstein and Katmadas that by flipping a few signs in the equations underlying BPS solutions one obtains non-supersymmetric ``almost-BPS'' solutions \cite{Goldstein:2008fq}, the authors together with Warner, Dall'Agata and Bobev have uncovered over a series of papers several very large classes of multicenter solutions in four-dimensional supergravity \cite{Bena:2009ev,Bena:2009en,Bena:2009fi,Bobev:2009kn,Bobev:2011kk}, that describe both non-supersymmetric black rings in Taub-NUT, as well as the seed solution for the most general under-rotating four-dimensional black hole. Furthermore, two of the authors and Dall'Agata have constructed the explicit duality transformations that allow one to start from an almost-BPS solution and obtain the most general class of multicenter extremal solutions of the STU model where one center is ``intrinsically'' non-supersymmetric and has four-dimensional rotation, while the other centers are ``locally-BPS'' \cite{Dall'Agata:2010dy}. Work is in progress by one of the authors and Bossard to obtain the most general multicenter solution with an arbitrary number of rotating non-BPS interacting centers \cite{bossard-ruef}.

Given the importance of finding new black rings, and given the power of the methods to construct multicenter non-BPS solutions, it is but natural to ask whether these methods can be used to construct black rings. After all, from the perspective of multicenter solutions \cite{Bates:2003vx},  a black ring in $\IR^4$ or Taub-NUT is nothing but the  five-dimensional supergravity uplift of a two-center solution, where one center has a nontrivial Gibbons-Hawking (or D6) charge but no horizon, while the other center is a D4-D2-D0 black hole. 

By now, three solutions for black rings in Taub-NUT have been constructed: the first one is supersymmetric \cite{Elvang:2005sa,Gaiotto:2005xt,Bena:2005ni}, and descends in four-dimensions to a BPS two-center solution where one of the centers is a D6 brane and the other center is a BPS D4-D2-D0 black hole. The second black ring solution  \cite{Bena:2009ev} is non-supersymmetric, and falls in the so-called almost-BPS class \cite{Goldstein:2008fq}: it is again a two-center solution, and each center is BPS by itself, but the orientation of the charge of one of the centers is reversed compared to the BPS solution, and hence supersymmetry is broken. The third solution is a solution of pure gravity, that does not have any electric or dipole magnetic charges \cite{Camps:2008hb}.

The black-ring-in-Taub-NUT solution that we construct in this paper does have electric and dipole magnetic charges, but is different from the BPS and almost-BPS solutions because the black-ring center descends in four dimensions to a {\it rotating} black hole, and hence is intrinsically non-supersymmetric. The uplifted solution therefore not only has angular momentum along the $S^1$ of the black ring horizon (as all black rings do) but also rotates along the $S^2$ of the horizon. In pure gravity such a black ring was constructed in five-dimensions by Pomeransky and Senkov 
\cite{Pomeransky:2006bd}, generalizing the non-rotating black ring of Emparan and Reall \cite{Emparan:2001wn}, but for rings with nontrivial dipole and electric charges no such generalization is known\footnote{For doubly-rotating black rings with only electric charges see for example \cite{Hoskisson:2008qq,Gal'tsov:2009da}.}.

The method for building our black ring solution uses the explicit four-dimensional S-duality map recently found in \cite{Dall'Agata:2010dy}, that allows one to take a multicenter solution with certain electric and magnetic charges and obtain, roughly speaking, a solution whose electric and magnetic charges are flipped. If one starts from a single-center solution for a four-dimensional rotating non-BPS black hole with one \Db{6} and three D2 charges, this duality map yields a rotating black hole with three D4 and one \Db{0} charge, which lifts up in M-theory to an M5-P rotating extremal black string. Furthermore, by turning on Wilson lines (axion vev's) in the original solution, one can also give M2 charges to this rotating black string. 

Having obtained a rotating black string solution, the question is whether one can bend it into a black ring and put in in Taub-NUT, and if possible in an asymptotically-five-dimensional solution. This is done by constructing a two-center solution, where one of the centers is the black string, and the other is a Gibbons-Hawking center (which descends in type IIA to a fluxed D6 
brane\footnote{By fluxed D6 brane we mean a D6 brane that has abelian worldvolume fluxes, and hence is T-dual to a single D3 brane at angles. A fluxed D6 brane preserves locally 16 supersymmetries, and uplifts to 11 dimensions to a smooth GH center \cite{Bena:2005va,Berglund:2005vb,Balasubramanian:2006gi}.}). Given that we obtained the black string by S-dualizing an \Db{6}-D2-D2-D2 black hole, we can ask what should be the object one needs to S-dualize to obtain a fluxed D6 brane. The answer is: any fluxed brane with D0 charge. 

Hence we propose to obtain a rotating charged non-BPS black ring in Taub-NUT by S-dualizing a two-center almost-BPS solution in which one of the centers is a rotating \Db{6}-D2-D2-D2 and the other center is a fluxed brane with D0 charge, or equivalently an M-theory object that rotates along the M-theory circle. The most obvious such object is a fluxed D4 brane, which uplifts in M-theory to a two-charge supertube wrapping the M-theory circle. Since the S-dual of a supertube is a fluxed D6 brane, the M-theory uplift of the S-dual configuration should produce a rotating black ring in Taub-NUT. Note that taking for the second center a fluxed D4 brane, one expects -- and this will indeed happen -- to have non trivial ``charges dissolved in fluxes'' in the final solution, coming from the second center.

By explicitly constructing the solution we find that the story is a bit more complicated: the interaction between the two centers induces after S-duality a nontrivial D6 charge at the black hole location; this causes its M-theory uplift to change topology, and not be a black ring but a KK black hole. However, there exists a very special choice of parameters that still maintains the black-ring horizon topology. For this choice, the solution describes a Taub-NUT rotating non-supersymmetric black ring with M5 dipole charges and M2 charges. 

The first obvious question is whether this Taub-NUT ring can be made into an asymptotically five-dimensional black ring. Indeed, by turning off the constant part in the Taub-NUT harmonic function, this space becomes $\IR^4$, and the solution for a BPS black ring in Taub-NUT \cite{Elvang:2005sa,Gaiotto:2005xt,Bena:2005ni} can be transformed straightforwardly into the solution for a BPS black ring in five dimensions \cite{Elvang:2004rt,Bena:2004de,Elvang:2004ds,Gauntlett:2004qy}. Nevertheless, for the black ring we construct this does not seem possible, at least after reasonable attempts. The reason is that the various constants that appear in the harmonic functions are already constrained by the requirement that the solution be a black ring, and one does not have the freedom to decompactify the Taub-NUT fiber at infinity and obtain a black ring in five dimensions.

The next question is whether our solution is the only Taub-NUT black ring within the class of solutions uncovered in \cite{Bena:2009ev,Bena:2009en,Bena:2009fi,Bobev:2009kn,Dall'Agata:2010dy}, or whether there are more black rings to be found. The solution we obtain falls into the class of extremal solution that can be obtained starting from a four-dimensional ``electrovac'' solution \cite{Bena:2009fi}. More precisely, our solution could in hind-sight have been constructed starting from a two-center Israel-Wilson base space \cite{Israel:1972vx}; however, the fact that one can construct a Taub-NUT black ring in this class of solutions is far from obvious, and the harmonic functions one must choose to get a black ring are far from being intuitive. One can also suspect that there may exist black ring solutions in the two new classes of solutions unveiled recently in \cite{Dall'Agata:2010dy}, and it would be clearly interesting to understand this.

Since the new black ring solution is obtained by acting with dualities on a known solution, this paper is organized as follows: we begin in section 2 by presenting both the starting solution -- a two-center configuration describing a \Db{6}-D2-D2-D2 black hole and a supertube -- and the dualities that we will perform on this solution. In section 3, we review the black string solution obtained in \cite{Dall'Agata:2010dy}, whose properties will be related to the local features of the black ring. The reader who is interested in the solution alone can skip directly to Section 4 where we give the explicit doubly spinning black ring in Taub-NUT solution, and explore its properties.

\section{The Starting Solution and the Explicit Duality Transformations}

As explained in the Introduction, we obtain our black ring solution in an indirect way : we start from a known two center solution, in Type IIA supegravity on $T^6$ and transform it using 6 T-dualities along this internal six-torus and an M-theory uplift in order to obtain the final solution. In this section, we therefore present first the seed solution, and then the T-duality rules that we will apply. 

\subsection{Starting solution} \label{starting}

Our starting solution is a two-center configuration corresponding to an a non-BPS black hole and a supertube in 11D supergravity on $T^6$ \cite{Bena:2009ev}, which when compactified to type IIA becomes
\bea
 ds_{10}^2 &=& -{I_4}^{-1/2} \, (dt + \omega)^2 + {I_4}^{1/2} \, ds_3^2 + \sum_{I=1}^3 \, \frac{{I_4}^{1/2}}{Z_I V} ds_{I}^2 \,, \nonumber \\ 
 \rm{e}^{2\Phi} &=& \frac{I_4^{3/2}}{Z^3 V^3} \,, \\
B^{(2)} &=& \sum_{I=1}^3 B^{(2)}_I dT_I \nonumber 
\eea
with 
\bea \label{B2start}
 B^{(2)}_1 = -{\mu\over Z_1} \,, \quad B^{(2)}_2 = -{\mu\over Z_2} \,, \quad B^{(2)}_3 = {K_3}-{\mu\over Z_3} \,,
\eea
for the NSNS-fields, where $Z=(Z_1 Z_2 Z_3)^{1/3}$ and where $ds^2_I$ and $dT_I$ are the metric and the volume form on each of the three 2-tori inside $T^6$. In detail, $ds^2_I=dy^2_{I,1}+dy^2_{I,2}$ and $dT_I=dy_{I,1}\wedge dy_{I,2}$, with $I=1,2,3$. The three-dimensional metric $ds_3^2$ is flat, and we have also defined
\bea
 I_4 = Z_1 Z_2 Z_3 V - \mu^2 V^2 \,.
\eea
In order to satisfy the equations of motions, $V$, $K_3$ and the $Z_I$'s need to be harmonic functions on the flat three-dimensional space, and $\mu$ and $\omega$ satisfy
\bea \label{eqmu}
 *_3 d \omega =  d( V \mu ) - V Z_3 d K_3 \,, \quad d *_3 d (V \mu) = d (V Z_3 ) *_3 d K_3 \,. 
\eea
The RR fields are 
\bea \label{Cstart}
C^{(1)} &=& A-{\mu V^2\over I_4}(dt+\omega) \,, \nonumber \\
C^{(3)} &=& \sum_{I=1}^3 C^{(3)}_I \wedge dT_I 
\eea
with 
\bea \label{C3start}
 C^{(3)}_1 \! = -{dt+\omega\over Z_1} -{\mu\over Z_1} A \,, \! \quad C^{(3)}_2 \! = -{dt+\omega\over Z_2} -{\mu\over Z_2} A \,, \! \quad C^{(3)}_3 \! = -{dt+\omega\over Z_3} + \Bigl({K_3}-{\mu\over Z_3}\Bigr) A + b_3 \,,
\eea
with
\bea \label{defvectors1}
*_3 d A &=&  -d V \,, \\ 
*_3 d b_3 &=&  K_3 d V - V d K_3 \,.
\eea
In order to perform the T-dualities, it will also be necessary to have the higher RR fields $C^{(5)}$ and $C^{(7)}$. They are given by the equations of motion of $C^{(1)}$ and $C^{(3)}$
\begin{eqnarray}
d C^{(5)} &=& -*_{10}\left( d C^{(3)} - d B^{(2)} \wedge C^{(1)} \right) + d B^{(2)} \wedge C^{(3)} \,,\\[2mm]\label{fieldstrengthbis}
d C^{(7)} &=& *_{10} d C^{(1)} + d B^{(2)}\wedge C^{(5)} \,.
\end{eqnarray}
Using the  explicit expressions for $C^{(1)}$, $C^{(3)}$ and $B^{(2)}$ (\ref{B2start}, \ref{Cstart}, \ref{C3start}), $C^{(5)} $ is explicitly given by 
\bea
C^{(5)} = C^{(5)}_{JK} \, dT_J \wedge dT_K 
\eea
with
\bea
C^{(5)}_{12} &=& {\mu\over Z_1 Z_2} (dt+\omega)- \gamma_3 + {\mu^2 \over Z_1 Z_2} A \,, \nonumber \\
C^{(5)}_{13} &=& {\mu\over Z_1 Z_3} (dt+\omega)- \gamma_2 - {\mu\over Z_1} \Bigl({K_3}-{\mu\over Z_3}\Bigr) A - {\mu \over Z_1} b_3 \,, \\ \nonumber
C^{(5)}_{23} &=& {\mu\over Z_2 Z_3} (dt+\omega)- \gamma_1 - {\mu\over Z_2} \Bigl({K_3}-{\mu\over Z_3}\Bigr) A - {\mu \over Z_2} b_3 \,.
\eea
Finally, $C^{(7)}$ is
\bea
C^{(7)} \!\!&=&\!\! \Bigl[ -{Z^3 V + \mu^2 V^2 \over V^2 Z^3} (dt+\omega)  - \beta + {\mu\over Z_I} \gamma_I - {\mu^3 \over Z^3} A + {\mu^2 \over Z_1 Z_2 } b_3 + K_3  \left( \! - \gamma_3 + { \mu^2 \over Z_1 Z_2 } A \right) \Bigr]  \nonumber \\ 
 && \quad \quad \quad \wedge dT_1 \wedge dT_2 \wedge dT_3\,,
\eea
with the vectors $\gamma_I$ and $\beta$ defined such that they verify the following relations
\bea \label{defvectors2}
*_3 d \gamma_I &=&  d Z_I \,, \\ 
*_3 d \beta &=&  Z_3 d K_3 - K_3 dZ_3 \,.
\eea
We finally have to specify what are exactly our harmonic functions, in order to describe the announced two-center solution. We parametrize the three-dimensional base space in cylindrical coordinates $(r,\theta,z)$ and take the black hole to be at the center of the space, at $r=0$, while the supertube is located along the positive $z$ axis at a distance $R$ from the origin. We denote the polar coordinates centered at the supertube position as
 $(\Sigma,\theta_\Sigma)$. Their relation to the polar coordinates $(r,\theta)$ centered at the origin is:
\be
\Sigma = \sqrt{r^2 + R^2 - 2 r R \cos\theta}\,,\qquad \cos\theta_\Sigma = {r\cos\theta-R\over \Sigma}\,.
\label{polarSigma}
\ee
The explicit functions are then  \cite{Bena:2009ev}

\bea \label{scalarsBHST}
 V &=& h + \frac{Q_6}{r} \,, \quad K_3 = k_3 + {d^{ST}_3 \over \Sigma} \,,  \nonumber  \\
 Z_1 &=& 1 + \frac{Q_1}{r} +\frac{Q^{ST}_1}{\Sigma} \,, \quad Z_2 = 1 + \frac{Q_2}{r} +\frac{Q^{ST}_2}{\Sigma} \,, \quad Z_3 = 1 + \frac{Q_3}{r} \,, \\ \nonumber 
 M &=& m_0 + \frac{m}{r} + \alpha \frac{\cos\theta}{r^2} + \frac{m^{ST}}{\Sigma} \,, \quad \mu = \frac{M}{V} + \frac{ d_3^{ST}}{2 \Sigma} + \frac{h \, Q_3 \, d_3^{ST}}{2 V \, r \, \Sigma} + \frac{Q_6 \, Q_3 \, d_3^{ST}}{V \, R \, r \, \Sigma} \cos\theta 
\eea
for the scalars and 
\bea \label{vectorsBHST}
 A \!\!&=&\!\! -Q_6 \cos\theta d\phi \,, \quad b_3 = \left( k_3 \, Q_6 \cos\theta - h \, d_3^{ST} \cos\theta_\Sigma - 
 Q_6 \, d_3^{ST} \, \frac{r - R \cos\theta}{R \, \Sigma} \right) d\phi \,, \nonumber \\ 
 \gamma_1 \!\!&=&\!\!  ( Q_1 \cos\theta + Q_1^{ST} \cos\theta_\Sigma ) d\phi \,, \ \gamma_2 =  ( Q_2 \cos\theta + Q_2^{ST} \cos\theta_\Sigma ) d\phi \,, \ \gamma_3  = Q_3 \cos\theta d\phi  \,, \nonumber  \\ 
 \omega \!\!&=&\!\! \Big( \kappa + m \cos\theta + m^{ST} \cos\theta_\Sigma  - \alpha \frac{\sin^2\theta}{r} - \frac{h \, d_3^{ST}}{2} \cos\theta_\Sigma \\ \nonumber 
&& \quad \quad \quad - ( Q_6 + h \, Q_3) \, d_3^{ST} \, \frac{r - R \cos\theta}{2 R \, \Sigma} - Q_6 \, Q_3 \, d_3^{ST} \, \frac{\sin^2\theta}{R \, \Sigma} \Big) d\phi \,, \\ \nonumber
 \beta \!\!&=&\!\!  \left( - k_3 \, Q_3 \cos\theta + d_3^{ST} \cos\theta_\Sigma + Q_3 \, d_3^{ST} \frac{r - R \cos\theta}{R \, \Sigma} \right) d\phi 
\eea
for the corresponding vector fields. Note that there is a constant term in the $K_3$ harmonic function that appears in $B^{(2)}_3$. This gives an additional contribution to the axion field at infinity. As we will see, this constant will be crucial after dualities in order to obtain a black ring.

\subsection{Performing 6 T-dualities}

We now review the general action of T-duality along all the $T^6$ directions $y_{1,1}, y_{1,2},\ldots y_{3,1},y_{3,2}$ \cite{Dall'Agata:2010dy} that we will apply on the solution presented in the previous subsection.

\subsubsection{NSNS fields}

The string metric and B-field have the general form
\be
ds^2_{10} = ds^2_4 + \sum_{I=1}^3 G_I  ds^2_I\,,\quad B^{(2)}=\sum_{I=1}^3 B_I dT_I\,.
\ee
Define the matrix
\be
E_I= \begin{pmatrix} G_I&B_I\\ -B_I& G_I\end{pmatrix} \,.
\ee
The sequence of two T-dualities along $y_{I,1}, y_{I,2}$ simply inverts the matrix $E_I$:
\be \label{NSNStsfo}
E_I\to \tilde E_I= E^{-1}_I = {1\over \Delta_I} \begin{pmatrix} G_I&-B_I\\ B_I& G_I\end{pmatrix}\,,\quad \Delta_I=G_I^2+B_I^2\,. 
\ee 
Hence the after T-duality on $T^6$ the torus metric and B-field are transformed as
\be
G_I\to \tilde G_I = {G_I\over \Delta_I}\,,\quad B_I = \tilde B_I = -{B_I\over \Delta_I}\,.
\ee
The dilaton transforms as
\be
e^{2\Phi}\to e^{2\tilde \Phi}= {e^{2\Phi}\over \Delta_1 \Delta_2 \Delta_3}\,.
\ee
%

\subsubsection{RR fields}

The action of T-duality on $T^6$ on the RR fields was obtained again in \cite{Dall'Agata:2010dy} by the brute force method of applying recursively the rules for T-duality along a compact direction. The rules we use are as follows. Let $y$ be the direction along which one performs T-duality and let us write the string metric, B-fields and RR gauge fields $C^{(p)}$ as
\bea
ds_S^{10} &=& G_{yy} (dy+A_\mu dx^\mu)^2 + \hat{g}_{\mu\nu} dx^\mu dx^\nu\,,\nn
B^{(2)}&=& B_{\mu y} dx^\mu \wedge (dy+A_\mu dx^\mu) + \hat{B}^{(2)}\,,\nn
C^{(p)}&=&C_y^{(p-1)}\wedge  (dy+A_\mu dx^\mu) + \hat{C}^{(p)}\,,
\label{trule0}
\eea
where the forms $\hat{B}^{(2)}$, $C_y^{(p-1)}$ and $\hat{C}^{(p)}$ are along the $x^\mu$ directions.
The T-duality transformed fields are
\bea
d{\tilde s}_{10}^2&=& G^{-1}_{yy} (dy-B_{\mu y} dx^\mu)^2 + \hat{g}_{\mu\nu} dx^\mu dx^\nu\,,\quad e^{2 \tilde \Phi}={e^{2\Phi}\over G_{yy}}\,,\nn
{\tilde B}^{(2)}&=& -A_{\mu} dx^\mu dy + \hat{B}^{(2)}\,,\\ \nonumber
{\tilde C}^{(p)}&=& \hat{C}^{(p-1)}\wedge(dy-B_{\mu y} dx^\mu)+C_y^{(p)}\,.
\label{trule}
\eea
The starting solution is 
\bea
&& ds^2_{10} = ds^2_4 + \sum_I G_I ds^2_I \,, \quad B^{(2)} = \sum_I B_I dT_I \,, \quad  C^{(1)} \,, \\
&& C^{(3)} = \sum_I C^{(3)}_I \wedge dT_I \,, \quad C^{(5)} = \sum_{I<J} C^{(5)}_{IJ} \wedge dT_I \wedge dT_J \,, \quad C^{(7)} = C^{(7)}_{123} \wedge dT_1 \wedge dT_2 \wedge dT_3 \,. \nonumber
\eea
One finds the fields after 6 T-dualities on $T^6$:
\bea \label{6Ttsfo}
&&d\tilde s^2_{10} = ds^2_4 +\sum_I {G_I\over \Delta_I} ds^2_I \,,\quad \tilde B^{(2)}=-\sum_I{ B_I\over \Delta_I} dT_I \,,\\
&&\tilde C^{(1)}=-C^{(7)}_{123}+{|\epsilon_{IJK}|\over 2}B_I C^{(5)}_{JK}-{|\epsilon_{IJK}|\over 2} B_I B_J C^{(3)}_K + B_1 B_2 B_3 C^{(1)}\,,\\
&&\tilde C^{(3)}_I = \Delta_I^{-1}\Bigl[B_I C^{(7)}_{123} -B_I |\epsilon_{IJK}| B_J C^{(5)}_{IK}+G_I^2 {|\epsilon_{IJK}|\over 2}C^{(5)}_{JK}\nonumber\\
&&\qquad\quad +B_1 B_2 B_3 C^{(3)}_I - G_I^2 |\epsilon_{IJK}| B_J C^{(3)}_K + G_I^2 {|\epsilon_{IJK}|\over 2} B_J B_K C^{(1)}\Bigr]\,,\\
&&\tilde C^{(5)}_{JK}=(\Delta_I\Delta_J)^{-1} \Bigl[-B_J B_K C^{(7)}_{123} + B_1 B_2 B_3 C^{(5)}_{JK} -|\epsilon_{IJK}| (G_I^2 B_J C^{(5)}_{JK}+G_J^2 B_I C^{(5)}_{IK})\nonumber\\
&&\qquad\quad +|\epsilon_{IJK}| B_K (G_I^2 B_J C^{(3)}_J+ G_J^2 B_I C^{(3)}_I)-G_I^2 G_J^2 |\epsilon_{IJK}| C^{(3)}_K + G_I^2 G_J^2 |\epsilon_{IJK}| B_K C^{(1)} \Bigr]\,,\\
&&\tilde C^{(7)}=(\Delta_I\Delta_J\Delta_K)^{-1}\Bigl[B_1 B_2 B_3 C^{(7)}_{123} + {|\epsilon_{IJK}|\over 2} (G_I^2 B_J B_K C^{(5)}_{JK}+ G_I^2 G_J^2 B_K C^{(3)}_K)+ G_1^2 G_2^2 G_3^2 C^{(1)}\Bigr]\,.\nonumber
\eea
%

\section{The Rotating Non-BPS Black String} \label{KKK=0}

Before presenting the main result of this paper: the new rotating non-BPS black ring in Taub-NUT, let us make a small detour to review the rotating non-BPS black string \cite{Dall'Agata:2010dy} which gives the infinite-ring or the near-ring limit of the black ring we will construct. This will allow us to explore its local properties, and understand the role of the Taub-NUT center in the final solution.

To obtain the black string solution, one has to start from the solution in section \ref{starting} with the the supertube fields turned off: $K_3=0$, $b_3=0$, $Q_1^{ST}=Q_2^{ST}=m^{ST}=0$. This simplifies the solution, which can now be written in a closed form in terms of five harmonic functions $V$, $M$ and $Z_I$, $I=1,2,3$. It describes the rotating \Db{6}-D2-D2-D2 black hole found in \cite{Bena:2009ev}:
\bea \label{BHfields}
 V \!\! &=&\!\! 1 + \frac{Q_6}{r} \,, \quad A = -Q_6 \cos\theta d\phi \,, \nonumber \\ 
 K_1 \!\! &=&\!\! K_2 = K_3 = 0 \,, \quad Z_I = L_I = 1 + \frac{Q_I}{r} \,, \\ \nonumber 
 M \!\! &=&\!\! m_0 + \alpha \frac{\cos\theta}{r^2} \,, \quad \mu = \frac{M}{V} \,, \quad \omega = m \cos\theta d\phi - \alpha \frac{\sin^2\theta}{r} d\phi \,.
\eea
The quartic invariant $I_4$ is given by
\bea
 I_4 = Z_1 Z_2 Z_3 V - M^2 \,.
\eea
After the 6 T-dualities, the new solution is given by
\bea \label{10DBSTR}
 ds_{10}^2 &=& -{I_4}^{-1/2} \, (dt + \omega)^2 + {I_4}^{1/2} \, ds_3^2 + \frac{{I_4}^{1/2}}{Z^3} \sum_{I=1}^3 \,  Z_I ds_{I}^2 \,, \nonumber \\ 
 \rm{e}^{2\Phi} &=& \frac{I_4^{3/2}}{Z^6} \,, \\
B^{(2)} &=& \frac{M}{Z^3} \sum_{I=1}^3 Z_I dT_I \,, \nonumber 
\eea
and 
\bea
C^{(1)} &=& { Z^3 \over I_4 } (dt+\omega) \,, \nonumber \\
C^{(3)} &=& -\sum_{I=1}^3 \gamma_I \wedge dT_I \,, \nonumber \\
C^{(5)}_{JK} &=& \frac{C_{IJK}}{Z_I} (dt+\omega) - \frac{M}{Z^3} \left( Z_J \gamma_K + Z_K \gamma_J \right) \,, \nonumber \\
C^{(7)} &=& \left( {2M \over Z^3} (dt+\omega) -\frac{M^2}{Z^3}\sum_I\frac{\gamma_I}{Z_I} + A \right) \wedge dT_1 \wedge dT_2 \wedge dT_3\,.
\eea

One can easily lift this solution back to eleven dimensions. We recall that this is done by 
\bea \label{lift11D}
 ds_{11}^2 &=& \rm{e}^{4\phi/3} ( d\psi + C^{(1)} )^2 + \rm{e}^{-2\phi/3} ds_{10}^2 \,, \\
 A^{(3)} &=& B^{(2)} \wedge d\psi + C^{(3)} \,. \nonumber
\eea
We then obtain
\bea \label{Bstrsol}
 ds_{11}^2 &=& \frac{2}{Z}(dt+\omega)d\psi + \frac{I_4}{Z^4}d\psi^2 + Z^2 ds_3^2 + \sum_I(\frac{Z_I}{Z} ds_I^2) \,, \\
 A^{(3)} &=& \sum_I \left( \frac{M}{Z^3} Z_I d\psi - \gamma_I \right) \wedge dT_I \,.
\eea
This solution corresponds to a non-supersymmetric black string. As found in \cite{Dall'Agata:2010dy}, the charges of the solution are \footnote{One may have naively expected that since $B^{(2)}$ has a non-zero value at infinity \eqref{10DBSTR}, this 
would affect the values of the D4, D2 and D0 charges, but this does not happen: the constant in the asymptotic B-field can be gauged away without affecting these charges; a more complete discussion of this can be found in \cite{Dall'Agata:2010dy}.}
:
\bea \label{chgesBSTR}
 Q^{D6} &=& \int_{S^2_{\infty}} d C^{(1)} = 0 \,, \nonumber \\
 Q^{D4}_{JK} &=& \int_{S^2_{\infty}\times T^2} d C^{(3)} = C_{IJK} Q_I \,, \\
 Q^{D2}_I &=& \int_{S^2_{\infty}\times T^4} d C^{(5)} = m_0 ( Q_J + Q_K ) \,, \nonumber \\ \nonumber
 Q^{D0} &=& \int_{S^2_{\infty}\times T^6} d C^{(7)} = Q_6 + m_0^2 ( Q_1 + Q_2 + Q_3 ) \,.
\eea
and the angular momentum of the solution in the transverse $\IR^3$ plane is 
\be
 J=\alpha \,.
\ee
Using \eqref{BHfields} and \eqref{Bstrsol}, the near-horizon geometry is:
\bea \label{NHstring}
 ds_5^2 \stackrel{r\to 0}{=} 2\frac{r}{Q} dt d\psi + Q^2 \frac{dr^2}{r^2} + \frac{Q^3 Q_6 - \alpha^2 \cos^2\theta}{Q^4} d\psi^2 - 2 \frac{\alpha}{Q} \sin^2\theta d\psi d\phi + Q^2  (d\theta^2+\sin^2\theta d\phi^2)\,,
\eea
where we defined $Q = (Q_1 Q_2 Q_3)^{1/3}$.

 By performing a coordinate change
\bea
\tilde\phi = \phi - \frac{\alpha}{Q^3} \psi \, ,
\label{tildephi}
\eea
one can bring the metric into the form
\bea
ds_5^2 = 2\frac{r}{Q} dt d\psi + \frac{Q^3 Q_6 - \alpha^2}{Q^4} d\psi^2  +
 Q^2 \frac{dr^2}{r^2} + Q^2 (d\theta^2+\sin^2\theta d\tilde\phi^2)\,.
\eea
The space spanned by $(t,\psi,r)$ is $AdS_3$ and thus the near-horizon geometry of this black string is a fibered $AdS_3\times S^2$ space. Note that for this black string solution we have not specified any periodicity for $\psi$, and hence the near-horizon metric can always be brought into the $AdS_3\times S^2$ form.

The induced metric on the horizon at $r=0$ is
\bea \label{hormetricBSTR}
 ds_{\rm{hor}}^2 &=& \frac{Q^3 Q_6 - \alpha^2 \cos^2\theta}{Q^4} d\psi^2 - 2 \frac{\alpha}{Q} \sin^2\theta d\psi d\phi + Q^2 d\Omega_2^2  \nonumber \\ 
 &=& Q^2 d \theta^2 + Q^2\frac{Q_6 Q^3 - \alpha^2}{Q_6 Q^3 - \alpha^2\cos^2\theta} \sin^2 \theta d \phi^2 \\ \nonumber 
&& + \frac{Q_6 Q^3 - \alpha^2\cos^2\theta}{Q^4}(d\psi -\frac{Q^3 \alpha}{Q_6 Q^3 - \alpha^2\cos^2\theta} \sin^2\theta d\phi)^2 \,.
\eea
The horizon has topology $S^2\times S^1$ and its area gives the entropy of the string:
\bea
S = 2 \pi \sqrt{Q_1 Q_2 Q_3 Q_6 - \alpha^2}\,.
\eea
%

\section{The Rotating non-BPS Black Ring in Taub-NUT} \label{BRinIW}

We now construct the full two-center solution, which describes a black ring in Taub-NUT. As we will discuss later this solution falls in the class of solutions with an Israel-Wilson base space, but because of its charges at infinity it is more appropriate to refer to it as a black ring in Taub-NUT than a black ring in an Israel-Wilson space. As hinted in the Introduction, we will see  explicitly that adding a second center with a nontrivial D6 charge bends the string of the previous section into non-trivial ring.

\subsection{The solution}

Starting from the solution constructed in section \eqref{starting} and applying six T-dualities on obtains the explicit type IIA two-center solution:
\bea
 ds_{10}^2 &=& -{I_4}^{-1/2} \, (dt + \omega)^2 + {I_4}^{1/2} \, ds_3^2 + \sum_{I=1}^3 \, \frac{{I_4}^{1/2}}{Z_I \Delta_I V} ds_{I}^2 \,, \nonumber \\ 
 \rm{e}^{2\Phi} &=& \frac{I_4^{3/2}}{Z^3 \Delta^3 V^3} \,, \\
B^{(2)} &=& \frac{1}{\Delta_1}{\mu\over Z_1}  \wedge dT_1 + \frac{1}{\Delta_2}{\mu\over Z_2}  \wedge dT_2 - \frac{1}{\Delta_3} \big(K_3 - {\mu\over Z_3} \big)  \wedge dT_3  \,, \nonumber 
\eea
where we recall that $\Delta_I= G_I^2+B_I^2$, explicitly given by
\bea
\Delta_1 = \frac{Z_2 Z_3}{V Z_1} \,, \quad \Delta_2 = \frac{Z_1 Z_3}{V Z_2} \,, \quad \Delta_3  = {Z_1 Z_2 \over V Z_3 } + K_3^2 - 2 {K_3 \mu \over Z_3 } \,,
\eea
and the functions $Z_I,K_3,\omega$ and $\mu$ are given in (\ref{scalarsBHST},\ref{vectorsBHST}). We also defined $\Delta = (\Delta_1 \Delta_2 \Delta_3)^{1/3}$. 

The corresponding RR-fields are 
\bea
C^{(1)} \!\!&=&\!\! \frac{1}{I_4}\left(Z^3 - V \mu Z_3 K_3 \right)(dt+\omega) + \beta \,, \nonumber \\ 
C^{(3)} \!\!&=&\!\! \sum_{I=1}^3 C^{(3)}_I \wedge dT_I \,, \\
C^{(5)} \!\!&=&\!\! \sum_{J,K=1}^3 C^{(5)}_{JK} \wedge dT_J \wedge dT_K \,, \\
C^{(7)} \!\!&=&\!\! C^{(7)}_{123} \wedge dT_1 \wedge dT_2 \wedge dT_3 
\eea
with\footnote{We have for convenience redefined $y_I^1$ to $-y_I^1$ for $I=1,2$. This changes the signs of $C^{(3)}_1$ and $C^{(3)}_2$.}
\bea
C^{(3)}_1 &=& -\frac{K_3}{Z_2} (dt + \omega ) + \gamma_1 - \frac{V \mu}{Z_2 Z_3} \, \beta \,, \quad C^{(3)}_2 = -\frac{K_3}{Z_1} (dt + \omega ) + \gamma_2 - \frac{V \mu}{Z_1 Z_3} \, \beta \,, \\
C^{(3)}_3 &=& - \frac{K_3}{V \Delta_3}(dt+\omega) -\gamma_3 -\frac{1}{\Delta_3}(K_3 -\frac{\mu}{Z_3})\, \beta \,, 
\eea
for $C^{(3)}$,
\bea
 C^{(5)}_{12} \!\!\!&=&\!\!\! \left( \frac{1}{Z_3} + \frac{K_3 V \mu}{Z^3} \right) (dt + \omega ) - b_3 - \frac{V \mu}{Z_3} \left( \frac{\gamma_1}{Z_1} + \frac{\gamma_2}{Z_2} \right)  + \frac{V^2\mu^2}{Z_1 Z_2 Z_3^2} \, \beta \,, \\ \nonumber
 C^{(5)}_{13} \!\!\!&=&\!\!\! \frac{1}{V Z_3 \Delta_3} \!\left(\! Z_1 - \frac{K_3 V \mu}{Z_2} \right) \! ( dt + \omega ) - b_2 - \frac{V \mu}{Z_2 Z_3} \gamma_3 + \frac{1}{\Delta_3} \left( K_3 - \frac{\mu}{Z_3} \right) \! \gamma_1 -\frac{V \mu}{Z_2 Z_3 \Delta_3} \left( K_3 - \frac{\mu}{Z_3} \right) \! \beta \,, \\ \nonumber
 C^{(5)}_{23} \!\!\!&=&\!\!\! \frac{1}{V Z_3 \Delta_3} \! \left( \! Z_2 - \frac{K_3 V \mu}{Z_1} \right) \! ( dt + \omega ) - b_1 - \frac{V \mu}{Z_1 Z_3} \gamma_3 + \frac{1}{\Delta_3} \left( K_3 - \frac{\mu}{Z_3} \right) \! \gamma_2 -\frac{V \mu}{Z_1 Z_3 \Delta_3} \left( K_3 - \frac{\mu}{Z_3} \right) \! \beta\,. 
\eea
for $C^{(5)}$ and 
\bea
C^{(7)}_{123} \!\!&=&\!\! \frac{1}{Z_3 \Delta_3} \left( \frac{2 \mu}{Z_3} - K_3 - \frac{K_3 V \mu^2}{Z^3} \right)(dt + \omega ) + A +\frac{1}{\Delta_3} \left( K_3 - \frac{\mu}{Z_3} \right) b_3 - \frac{V^2 \mu^2}{Z_1 Z_2 Z_3^2} \, \gamma_3  \nonumber \\ 
 && \!\! + \frac{V \mu}{\Delta_3 Z_3} \left( K_3 - \frac{\mu}{Z_3} \right) \left( \frac{\gamma_1}{Z_1} + \frac{\gamma_2}{Z_2} \right)  - \frac{V^2 \mu^2}{Z_1 Z_2 Z_3^2 \Delta_3} \left( K_3 - \frac{\mu}{Z_3} \right) \beta 
\eea
for $C^{(7)}$.

\paragraph{Reconstructing the 11D fields.}

One can lift this solution to eleven dimensions, using \eqref{lift11D}. This gives
\bea \label{new11Dmet}
ds_{11}^2 &=& -\widetilde{Z}^{-2}\left(dt+\tilde{k}\right)^2+\widetilde{Z} ds^2_4+\sum_{I=1}^3{\widetilde{Z}\over \widetilde{Z}_I} ds^2_I\,, \\ \nonumber
 A^{(3)} &=& \sum_I \Bigl[ -\frac{1}{\widetilde{Z}_I} (dt+\tilde{k}) + \tilde{a}_I \Bigr] \wedge dT_I \,,\eea
with 
\bea \label{newmetricdatas}
 \widetilde{Z}_1 &=& \frac{Z_2}{K_3} \,, \quad \widetilde{Z}_2 = \frac{Z_1}{K_3} \,, \quad \widetilde{Z}_3 = \frac{Z_1 Z_2}{Z_3 K_3} + V (K_3 -2 \frac{\mu}{Z_3}) \,, \\ \nonumber
 \widetilde{k} &=&  \widetilde{\mu} (d\psi+\widetilde{A}) + \omega\,,\quad \widetilde{\mu} = -\frac{Z^3}{(Z_3 K_3)^2} + \frac{V \mu}{Z_3 K_3} \,,
\eea
and
\bea \label{newggedatas}
\tilde{a}_I &=& \widetilde{K}_I (d\psi + \widetilde{A}) + \tilde{b}_I \,,\\ \nonumber
 \widetilde{K}_1 &=& -\frac{Z_1}{K_3 Z_3} \,, \quad \widetilde{K}_2 = -\frac{Z_2}{K_3 Z_3} \,, \quad \widetilde{K}_3 = -\frac{1}{K_3} \,, \\ \nonumber
\tilde{b}_1 &=& \gamma_1 \,, \quad \tilde{b}_2 = \gamma_2 \,, \quad \tilde{b}_3 = -\gamma_3\,.
\eea
The four-dimensional base is given by
\bea
 ds^2_4 &=& \widetilde{V}^{-1}(d\psi+\widetilde{A})^2+ \widetilde{V} ds^2_3\,, \\ \nonumber
\widetilde{V} &=& \widetilde{V}_+\widetilde{V}_- \,, \quad \widetilde{A}=\beta  \,,
\eea
where we relabeled 
\be
\widetilde{V}_+=K_3 \quad \mathrm{and}\quad \widetilde{V}_-=Z_3\,.
\ee 
With these redefinitions, one can see that equation \eqref{defvectors2} can be rewritten as
\be
*d\beta = \widetilde{V}_- d\widetilde{V}_+ - \widetilde{V}_+ d\widetilde{V}_- \,.
\ee
Hence, the base space of our solution is an Israel-Wilson space \cite{Israel:1972vx}. One should recall that in  \cite{Bena:2009fi} the authors and Warner have shown that one can construct a certain class of eleven-dimensional non-supersymmetric solutions starting from any four-dimensional Euclidean electrovac solution, and that moreover when this electrovac solution is an Israel-Wilson space the solution can be obtained by solving in a linear algorithm certain harmonic equations in $\IR^3$.  It would be of course interesting to disentangle the building blocks of this black ring solution in the language of solutions with an Israel-Wilson base, and to understand what is the physics of the more general solutions one can construct with these building blocks.

\subsection{Physical properties}

We now turn to the analysis of the physical properties of the constructed solution. 

\paragraph{Regularity.} 

For absence of Dirac-Misner strings $\omega$ has to vanish at $\theta=0$ and $\theta=\pi$, which for a two-center solution  imposes three constraints on the parameters, one for $\theta=\pi$, one for $\theta=0$ and $r<R$ and the last one for $\theta=0$ and $r>R$. These constraints can be solved by taking
\bea
 \kappa = -m = \frac{h \, Q_3 + Q_6}{2 R} \, d_3^{ST} \,, \quad m^{ST} = \frac{ h \, Q_3 + Q_6 + h \, R }{2 R} \, d_3^{ST} \,.
 \label{reg1}
\eea
We also want to make sure that the pole at $\Sigma=0$ is a Taub-NUT center, and therefore regular, and locally $\mathbb{R}^4$. At this location, $\widetilde{Z}_1$ and $\widetilde{Z}_2$ are constant, but $\widetilde{Z}_3$ has a pole
\bea
 \widetilde{Z}_3 = \frac{Q_1^{ST} Q_2^{ST} R^2 - (Q_6 + h R) (Q_3 + R) {d_3^{ST}}^2 }{d_3^{ST} R (Q_3 + R) \Sigma} + O(1) \,
\eea
and one therefore has to impose
\bea
 {d_3^{ST}}^2 = \frac{Q_1^{ST} Q_2^{ST} R^2}{(Q_6 + h R) (Q_3 + R)} 
\label{reg2}
\eea
for this pole to vanish. With these values for $m^{ST}$ and $d_3^{ST}$, one can then check that $\tilde{\mu}$ does vanish for $\Sigma \to 0$, and this ensures the required regularity. Finally, it is easy to check that $\widetilde{V} \sim 1/\Sigma$, near $\Sigma=0$, and hence the four-dimensional space looks locally like $\IR^4$. 

\paragraph{Asymptotic behavior.} 

Asymptotically, the eleven dimensional metric becomes 
\bea
 ds_{\infty}^2 &=& C^{1/3} \Bigl[ g_{tt\infty} \left( dt + \mu_{\infty} (d\psi + A_\infty ) \right) ^2 \\ \nonumber
 && \quad \quad \quad + (g_{\psi\psi\infty} - g_{tt\infty}\mu_{\infty}^2)(d\psi+A_\infty)^2 + ds_3^2 + \sum_I g_{I\infty} ds_I^2 \Bigr]
\eea
with 
\bea
 C &=& 1 + h k_3^2 - 2 k_3 m_0 \,, \quad g_{tt\infty} = -\frac{k_3^2}{C} \,, \nonumber \\ 
 g_{\psi\psi\infty} - g_{tt\infty}\mu_{\infty}^2 &=& \frac{1}{k_3^2} \,, \quad \mu_\infty = \frac{-1 + k_3 m_0}{k_3^2} \,, \quad A_\infty = \big(\frac{d_3^{ST} Q_3}{R} + (d_3^{ST} - k_3 Q_3) \cos\theta \big) d\phi \,, \nonumber \\ 
 g_{1\infty} &=& 1 \,, \quad g_{2\infty} = 1 \,, \quad g_{3\infty} = {1 \over C } \,.
\eea
Let us perform the following change of coordinates
\bea
  \psi & \to & \psi - \frac{g_{tt\infty}\mu_{\infty}}{g_{\psi\psi\infty}} \, t \,, \quad t \to C^{-1/6} \Big (-\frac{g_{tt\infty} (g_{\psi\psi\infty} - g_{tt\infty}\mu_{\infty}^2)}{g_{\psi\psi\infty}}\Big )^{-1/2} \, t \,, \nonumber \\
 r & \to & C^{-1/6} r \,, \quad y_{I,a} \to  C^{-1/6} (g_{I\infty})^{-1/2} \, y_{I,a} \,, \ I=1,2,3 \ a=1,2 \,.  
\eea
Then the asymptotic metric becomes
\bea
 ds_{\infty}^2 = -dt^2 + ds_3^2 + \sum_I ds_I^2 + C^{1/3} g_{\psi\psi\infty}(d\psi+ (d_3^{ST} - k_3 Q_3)\cos\theta d\phi )^2
\eea
and hence the solution is not asymptotically $\IR^{4,1}$ but $\IR^{3,1} \times S^1$. More precisely, because of the non-trivial fibration, the space is asymptotically Taub-NUT. It is regular under the following identifications
\bea \label{asymptangles}
  (\phi,\psi) \sim \Big (\phi + 2 \pi ,\psi - 2\pi \Big ( { d_3^{ST} Q_3 \over R } + d_3^{ST} - k_3 Q_3\Big )\Big ) \sim \Big (\phi + 2\pi ,\psi - 2\pi \Big ( { d_3^{ST} Q_3 \over R }-d_3^{ST} + k_3 Q_3\Big ) \Big)\,.\nonumber\\
\eea

\paragraph{Charges at infinity.}
 
 The solution dimensionally reduced to four-dimensions is asymptotically flat, and its Einstein-frame metric is
 \bea
 ds_{E}^2 &=& -{I_4}^{-1/2} \, (dt + \omega)^2 + {I_4}^{1/2} \, ds_3^2 \,.
\eea
The corresponding asymptotic charges are the mass
\bea
G_4 M = {Q_6 + h \,(Q_1 + Q_1^{ST} + Q_2 + Q_2^{ST} + Q_3 - 2 m_0 \, d_3^{ST} ) \over 4 (h - m_0^2)^{3/4}}\,,
\eea
(with $G_4$ the 4D Newton's constant), the angular momentum
\bea
G_4 J = \alpha + {d_3^{ST}\over 2} \Bigl[Q_6 + Q_3 \Bigl( h + 2 {Q_6\over R} \Bigr) \Bigr] \,,
\eea
the D6 charge
\bea \label{D6ch}
Q^{D6}= d_3^{ST} - k_3\, Q_3 \,,
\eea
the three D4 charges
\bea \label{D4ch}
&& Q^{D4}_{23} = Q_1+Q_1^{ST}- m_0 Q^{D6} \,, \quad Q^{D4}_{13} = Q_2 + Q_2^{ST} - m_0 Q^{D6} \,, \\
&& Q^{D4}_{12} = {(k_3 m_0 - 1) Q_3 + d_3^{ST} (m_0 - h\,k_3 )\over C } = -Q_3 - {h\,k_3 - m_0 \over C} Q^{D6} \,, \nonumber
\eea
the three D2 charges
\bea \label{D2ch}
Q^{D2}_1 &=& {h\,k_3 - m_0 \over C} Q^{D4}_{13}- m_0 Q_3 \,, \nonumber\\
Q^{D2}_2 &=& {h\,k_3 - m_0 \over C} Q^{D4}_{23}- m_0 Q_3 \,, \nonumber\\
Q^{D2}_3 &=& h \, d_3^{ST} - k_3 Q_6 - m_0 Q^{D4}_{13} - m_0 Q^{D4}_{23} - m_0^2 Q^{D6}\,,
\eea
and the D0 charge
\bea \label{D0ch}
Q^{D0} = -Q_6 - {h\,k_3 - m_0\over C } Q^{D2}_3 - m_0^2 Q_3 \,.
\eea
We can see that these charges are quite different from those of the lonely black string \eqref{chgesBSTR}. This of course comes from the presence of the second center, which is a fluxed D6-brane with nontrivial D4, D2 and D0 charges. Furthermore, the M-theory uplift has nontrivial magnetic fields sourced by M5 branes, which give rise to extra M2 charges dissolved in fluxes and to extra angular momentum. 

\paragraph{Topology of the horizon.}

In order for this solution to describe a black ring at $r=0$ it is necessary that the $\psi$-fiber does not shrink, but stays finite. Since the size of this fiber is 
\bea
 (\widetilde{V}_+\widetilde{V}_-)^{-1} = \big(l_3 + {Q_3 \over r} \big)^{-1} \big(k_3 + { d_3^{ST} \over \Sigma } \big)^{-1} \,,
\eea
it generically shrinks to zero as $r \to 0$, and then the solution does not describe a black ring, but a KK-black hole. Indeed, this is the generic behavior one expects for a solution constructed with an Israel-Wilson base space, and one could naively think that one cannot construct a black ring in this class of solutions. This is not so: the effective D6 charge at $r=0$ is
\be
 Q^{\rm{D}6}_{\rm{eff}} = \big(k_3 + { d_3^{ST} \over R } \big) Q_3 
\ee
and, because we have allowed the constant $k_3$ to be non-zero, we can tune it to cancel this effective D6 charge: 
\be \label{k3value}
 k_3 = - \frac{d_3^{ST}}{R} \,.
\ee
For this particular value of $k_3$, the size of the fiber stays finite, and we expect to have a black ring. Indeed, the horizon metric becomes
\bea
 ds_{\rm{hor}}^2 \!\! &=& \!\! Q^2 d \theta^2 + Q^2\frac{Q_6 Q^3 - \alpha^2}{Q_6 Q^3 - \alpha^2\cos^2\theta} \sin^2 \theta d \phi^2 \\ \nonumber 
&& \quad \quad + \frac{Q_6 Q^3 - \alpha^2\cos^2\theta}{Q^4} \left( d\psi -\big( d_3^{ST} + \frac{Q^3 \alpha}{Q_6 Q^3 -  \alpha^2\cos^2\theta} \sin^2\theta \big) d\phi \right)^2 \\ \nonumber
\!\! &=& \!\! Q^2 d \theta^2 \! + Q^2 \sin^2 \theta d \phi^2 \! + \frac{Q_6 Q^3 - \alpha^2\cos^2\theta}{Q^4} (d\psi - d_3^{ST} d\phi )^2 \, - 2\frac{\alpha}{Q}\sin^2\theta (d\psi - d_3^{ST} d \phi ) d\phi
\eea
with $Q=(Q_1 Q_2 Q_3)^{2/3}$. After the redefinition 
\be \label{redefphi}
\tilde{\psi}= \psi - d_3^{ST} \phi \,, 
\ee
this is exactly the metric \eqref{hormetricBSTR} that we obtained for the black string, and thus has topology $S^2\times S^1$. Here, regularity 
implies the identifications
\bea\label{psitildeangle}
 (\phi, \tilde{\psi}) \sim  (\phi + 2\pi , \tilde{\psi} ) \,.
 \label{ident}
\eea
Note that, using the constraint (\ref{k3value}), the second of the identifications  \eqref{asymptangles}  obtained from the asymptotic analysis becomes
\be
(\phi,\psi)  \sim (\phi + 2\pi ,\psi + 2\pi d_3^{ST} )\,,
\ee
which, given the definition of $\tilde{\psi}$, is exactly equivalent to the identification \eqref{psitildeangle}.

It is important to note that the condition \eqref{k3value} is very particular from the perspective of solutions with an Israel-Wilson base space. This base space can be ambipolar\footnote{It can have regions where the signature alternates from $(+,+,+,+)$ to $(-,-,-,-)$.} and still yield a physical five-dimensional solution (much like Gibbons-Hawking spaces do \cite{Giusto:2004kj,Bena:2005va,Berglund:2005vb}). The choice  \eqref{k3value} places the black ring exactly on top of the boundary between regions of opposite signature. While one could be worried that regularity can be affected by this very particular choice, it is easy to see that all the previously-required conditions are still respected, and therefore the solution is perfectly fine. 

The entropy of the black ring is still given by
\bea
 S = 2 \pi \sqrt{Q_1 Q_2 Q_3 Q_6 - \alpha^2} \,.
\eea
As we already discussed, because of the charges dissolved in fluxes, coming from the second center, one does not expect this entropy to be given by the quartic invariant of the asymptotic charges \eqref{D6ch}-\eqref{D0ch}. For all that it's worth, the latter is
\be
 J_4(Q^{D0},Q^{D2}_I,Q^{D4}_{JK},Q^{D6}) = - (Q_1 + Q_1^{ST}) (Q_2 + Q_2^{ST}) Q_3 Q_6 - \frac{1}{4}{d_3^{ST}}^2 (Q_6 - h Q_3)^2\,.
\ee
It is interesting to note that this quartic invariant is relatively simple, and does not involve the constants $k_3$ and $m_0$, despite the fact that they do appear in the charges. Moreover, since the charges dissolved in flux come from the presence of the second center, by turning off its charges, $Q_1^{ST}$, $Q_2^{ST}$ and $d_3^{ST}$, one recovers the usual entropy formula $S =2 \pi \sqrt{J_4(Q^{D0},Q^{D2}_I,Q^{D4}_{JK},Q^{D6})-\alpha^2}$.

Having obtained a rotating black ring in Taub-NUT, it is natural to ask whether one can choose solution parameters such that the Taub-NUT fiber decompactifies at infinity, and the solution describes a black ring in $\IR^{4,1}$. While this can be done for a generic choice of parameters, when one imposes that the topology of the horizon be $S^2 \times S^1$ \eqref{k3value}, this restricts the number of parameters, and the asymptotically $\IR^{4,1}$ solution disappears. Hence, one can obtain a two-center five-dimensional solution with a rotating black hole, but not a five-dimensional black ring. It would be interesting to understand the physical intuition behind this.

\paragraph{Near horizon geometry.}

This is an interesting question to investigate, because one would naively expect that a black string ring placed right on the boundary between regions of the base of opposite signatures would have an unusual near-horizon metric. Amazingly enough, this does not happen. When $k_3 = - d_3^{ST} / R$ and $r\to 0$ 
\bea
 Z^3 \Delta^3 V^3 &\stackrel{r \to 0}{\sim}& \frac{(Q_1 Q_2 Q_3)^2}{r^6} \,, \quad I_4 \sim \frac{Q^3 Q_6 - \alpha^2 \cos^2\theta}{r^4} \,, \nonumber \\
 \beta &\sim & -d_3^{ST} d\phi \,,  \quad \omega \sim - \alpha \frac{\sin^2\theta}{r} d\phi \,, \quad  \widetilde{C}^{(1)}_t \sim \frac{Q^3}{Q^3 Q_6 - \alpha^2 \cos^2\theta}r \,.
\eea
The near horizon geometry is 
\bea
 ds_5^2 = 2\frac{r}{Q} dt d\tilde\psi + \frac{Q^3 Q_6 - \alpha^2 \cos^2\theta}{Q^4} d\tilde\psi^2 - 2 \frac{\alpha}{Q} \sin^2\theta d\tilde\psi d\phi + Q^2 \frac{dr^2}{r^2} + Q^2 (d\theta^2+\sin^2\theta d\phi^2)\,,
 \label{nearring}
\eea
with $\tilde\psi$ defined in \eqref{redefphi}. After this redefinition of $\psi$, the near-horizon geometry of the black ring coincides with that of the black string given in \eqref{NHstring}. Much like for the black string, the coordinate redefinition
\bea
\tilde\phi = \phi - \frac{\alpha}{Q^3} \tilde \psi
\label{tildephi1}
\eea
brings the metric into the $AdS_3\times S^2$ form
\bea
ds_5^2 = 2\frac{r}{Q} dt d\tilde\psi + \frac{Q^3 Q_6 - \alpha^2}{Q^4} d\tilde\psi^2  +
 Q^2 \frac{dr^2}{r^2} + Q^2 (d\theta^2+\sin^2\theta d\tilde\phi^2)\,.
 \label{nicemetric}
\eea

Note that the coordinate redefinition (\ref{tildephi1}) is compatible with the periodic identification of the angles. Indeed, in (\ref{ident}), the angle $\tilde{\psi}$ is kept fixed. Therefore, it directly tranlates into
\be
 (\tilde{\phi},\tilde{\psi}) \sim (\tilde{\phi} + 2 \pi , \tilde{\psi} ) \,,
\ee
which is exactly what is needed for the regularity of the metric  \eqref{nicemetric}.

\paragraph{The thin ring limit}

The non-BPS black ring solution constructed in this section admits a thin ring limit, in which the solution appears as a deformation of the non-BPS black string of section \ref{KKK=0}. As it is clear from the construction of the solution, the black ring is expected to reduce to the black string when the distance $R$ between the two centers becomes much larger than all the other scales. In the following we verify this expectation. 

In the large $R$ limit one has
\be
d_3^{ST}\approx \sqrt{Q_1^{ST} Q_2^{ST}}\,,\quad k=-m\approx  \frac{h \, Q_3 + Q_6}{2 R} \,  \sqrt{Q_1^{ST} Q_2^{ST}}\,, \quad m^{ST} \approx \frac{ h}{2} \,  \sqrt{Q_1^{ST} Q_2^{ST}}\,.
\ee
With the choice (\ref{k3value}), the function $K_3$ vanishes in the large $R$ limit:
\be
K_3 =d_3^{ST}\Bigl(\frac{1}{\Sigma}-\frac{1}{R}\Bigr)\approx \frac{\sqrt{Q_1^{ST} Q_2^{ST}}}{R^2} \,r\cos\theta\,.
\ee
This causes the various metric coefficients to either vanish or diverge in the limit:
\bea
&&\tilde V = K_3 Z_3 \,,\quad \tilde A = \beta \approx -\sqrt{Q_1^{ST} Q_2^{ST}} d\phi\,,\nonumber\\
&&\tilde Z_1 = \frac{Z_2}{K_3}\,,\quad \tilde Z_2 = \frac{Z_1}{K_3}\,,\quad \tilde Z_3 \approx \frac{Z_1 Z_2}{Z_3}\frac{1}{K_3}\,,\\
&&\tilde \mu \approx - \frac{Z_1 Z_2}{Z_3}\frac{1}{K_3^2}\,,\quad \tilde \omega = \omega\,.
\eea
The 11D metric, however, has a good limit, given by\footnote{The coefficient of $d\tilde\psi^2$ follows from the identity
\be
\tilde Z \tilde V^{-1}-\tilde Z^{-2} \tilde \mu^2\approx \frac{I_4}{Z^4}\,. \nonumber
\ee}
\be
ds^2_{11}\approx \frac{2}{Z} (dt+\omega)d\tilde\psi +  \frac{I_4}{Z^4} d\tilde \psi^2+ Z^2 ds^2_3 +\sum_I \frac{Z_I}{Z}\,,
\ee
where 
\be
\tilde\psi=\psi-\sqrt{Q_1^{ST} Q_2^{ST}} d\phi\,.
\ee
As expected, this coincides with the black string metric (\ref{Bstrsol}). The terms of higher order in the $1/R$ expansion of the black ring metric represent the deformation of the black string into a 
thin black ring, and are thus within the range of validity of the blackfold approximation. The balancing condition for the rotating black ring does not appear explicitly in the equations above, but is contained implicitly in the regularity conditions (\ref{reg1}, \ref{reg2}), which are solved by tweaking the parameters of the horizonless center.

\paragraph{Local or not local ?}

We have seen in this paper that when one transforms a two-center solution under dualities, the local properties of the solution near one of the centers are influenced by the presence of the other center. Thus, since applying 
6 T-dualities on a lonely spinning black hole always yields a spinning black string with horizon topology $S^2 \times S^1$, one may have naively expected that if one starts from a two-center solution, the black hole always transforms into an object that looks locally like a spinning black string. Our explicit calculation has shown that this only happens for a very specific choice of parameters. 

The reason for this is that the solution with a black hole and a supertube also has a nontrivial Wilson line (more precisely a 3-form potential with two legs on a $T^2$ and one along the Taub-NUT fiber, coming from a nonzero harmonic function $K_3$). If one applies 6 T-dualities on a single-center black hole in a background with a generic Wilson line, the resulting solution will not be a black string. The horizon topology only becomes 
$S^2 \times S^1$ when the Wilson line in the original geometry is absent: this is the black string solution presented in Section 3. For the two-center solution the condition for obtaining a black ring, \eqref{k3value}, can be traced back to the vanishing of the Wilson line at the black hole location in the original geometry. The near-ring solution then becomes exactly that of the black string. Note moreover that the topology change produced by the presence of a Wilson loop at the horizon does not affect the explicit value of the entropy.

\section{Conclusion}

In this paper we presented a new non-BPS doubly-spinning black ring solution in Taub-NUT. This is the first black ring solution that has both dipole charges and rotation along the $S^2$ of the horizon. This solution has been obtained by acting with duality transformations on a two-center configuration corresponding to a black hole and a supertube. The supertube becomes after dualities a fluxed D6-brane, and permits to bend the second center into a black ring. This construction has some noticeable properties that are worth recalling. 

First of all, we have discovered that the presence of a tiny Wilson line near the black hole horizon alters completely the topology of the solution obtained after six T-dualities. Second, the final solution belongs to the class of non-BPS solutions constructed starting from an Israel-Wilson space, in which one did not naively expect to find black ring solutions; furthermore, the solution is regular despite the fact that from the perspective of the Israel-Wilson base space the black ring center is located exactly at the boundary of two regions with different signature. Third our method for constructing the solution does not allow to decompactify the Taub-NUT space to asymptotically $\IR^4$, and to obtain a rotating black ring in five dimensions. It would be interesting to see if this could be done by dualizing other almost-BPS two-center solutions, where the fluxed brane is not a D4 brane but a D2 or D0 brane, or by starting with an Israel-Wilson solution where the fluxed brane is a D6 brane.  It is also interesting to understand how to add angular momentum in solutions that have an Israel-Wilson base space, or in the more general solutions found in \cite{Dall'Agata:2010dy}.

\bigskip
\noindent {\bf Acknowledgments:} We would like to thank Guillaume Bossard, Gianguido Dall'Agata, Maria Rodriguez, Amitabh Virmani, and Nick Warner for interesting discussions. The work of I.~B. is supported by the DSM CEA/Saclay, the ANR grant 08--JCJC--0001--0, the ERC Starting Independent Researcher Grant 240210 -- String--QCD--BH.

\smallskip
%

\bibliographystyle{utphys}
\bibliography{micro}

\providecommand{\href}[2]{#2}\begingroup\raggedright\begin{thebibliography}{10}

\bibitem{Emparan:2001wn}
R.~Emparan and H.~S. Reall, ``{A Rotating black ring solution in
  five-dimensions},''
  \href{http://dx.doi.org/10.1103/PhysRevLett.88.101101}{{\em Phys.Rev.Lett.}
  {\bf 88} (2002)  101101}, \href{http://arxiv.org/abs/hep-th/0110260}{{\tt
  arXiv:hep-th/0110260 [hep-th]}}.

\bibitem{Emparan:2004wy}
R.~Emparan, ``{Rotating circular strings, and infinite nonuniqueness of black
  rings},'' \href{http://dx.doi.org/10.1088/1126-6708/2004/03/064}{{\em JHEP}
  {\bf 0403} (2004)  064}, \href{http://arxiv.org/abs/hep-th/0402149}{{\tt
  arXiv:hep-th/0402149 [hep-th]}}. Erratum added online, May/18/2006.

\bibitem{Elvang:2004rt}
H.~Elvang, R.~Emparan, D.~Mateos, and H.~S. Reall, ``{A Supersymmetric black
  ring},'' \href{http://dx.doi.org/10.1103/PhysRevLett.93.211302}{{\em
  Phys.Rev.Lett.} {\bf 93} (2004)  211302},
  \href{http://arxiv.org/abs/hep-th/0407065}{{\tt arXiv:hep-th/0407065
  [hep-th]}}.

\bibitem{Bena:2004de}
I.~Bena and N.~P. Warner, ``{One ring to rule them all ... and in the darkness
  bind them?},'' {\em Adv. Theor. Math. Phys.} {\bf 9} (2005)  667--701,
\href{http://arxiv.org/abs/hep-th/0408106}{{\tt arXiv:hep-th/0408106}}.

\bibitem{Elvang:2004ds}
H.~Elvang, R.~Emparan, D.~Mateos, and H.~S. Reall, ``{Supersymmetric black
  rings and three-charge supertubes},''
  \href{http://dx.doi.org/10.1103/PhysRevD.71.024033}{{\em Phys.Rev.} {\bf D71}
  (2005)  024033}, \href{http://arxiv.org/abs/hep-th/0408120}{{\tt
  arXiv:hep-th/0408120 [hep-th]}}.

\bibitem{Gauntlett:2004qy}
J.~P. Gauntlett and J.~B. Gutowski, ``General concentric black rings,'' {\em
  Phys. Rev.} {\bf D71} (2005)  045002,
\href{http://arxiv.org/abs/hep-th/0408122}{{\tt hep-th/0408122}}.

\bibitem{Elvang:2007hg}
H.~Elvang, R.~Emparan, and P.~Figueras, ``{Phases of five-dimensional black
  holes},'' \href{http://dx.doi.org/10.1088/1126-6708/2007/05/056}{{\em JHEP}
  {\bf 0705} (2007)  056}, \href{http://arxiv.org/abs/hep-th/0702111}{{\tt
  arXiv:hep-th/0702111 [hep-th]}}.

\bibitem{Emparan:2007wm}
R.~Emparan, T.~Harmark, V.~Niarchos, N.~A. Obers, and M.~J. Rodriguez, ``{The
  Phase Structure of Higher-Dimensional Black Rings and Black Holes},''
  \href{http://dx.doi.org/10.1088/1126-6708/2007/10/110}{{\em JHEP} {\bf 0710}
  (2007)  110}, \href{http://arxiv.org/abs/0708.2181}{{\tt arXiv:0708.2181
  [hep-th]}}.

\bibitem{Emparan:2009at}
R.~Emparan, T.~Harmark, V.~Niarchos, and N.~A. Obers, ``{Essentials of
  Blackfold Dynamics},'' \href{http://dx.doi.org/10.1007/JHEP03(2010)063}{{\em
  JHEP} {\bf 1003} (2010)  063}, \href{http://arxiv.org/abs/0910.1601}{{\tt
  arXiv:0910.1601 [hep-th]}}.

\bibitem{Emparan:2009zz}
R.~Emparan, T.~Harmark, V.~Niarchos, and N.~A. Obers, ``{Blackfold approach for
  higher-dimensional black holes},'' {\em Acta Phys.Polon.} {\bf B40} (2009)
  3459--3506.

\bibitem{Grignani:2011mr}
G.~Grignani, T.~Harmark, A.~Marini, N.~A. Obers, and M.~Orselli,
  ``{Thermodynamics of the hot BIon},''
  \href{http://arxiv.org/abs/1101.1297}{{\tt arXiv:1101.1297 [hep-th]}}.

\bibitem{Goldstein:2008fq}
K.~Goldstein and S.~Katmadas, ``{Almost BPS black holes},''
  \href{http://dx.doi.org/10.1088/1126-6708/2009/05/058}{{\em JHEP} {\bf 0905}
  (2009)  058}, \href{http://arxiv.org/abs/0812.4183}{{\tt arXiv:0812.4183
  [hep-th]}}.

\bibitem{Bena:2009ev}
I.~Bena, G.~Dall'Agata, S.~Giusto, C.~Ruef, and N.~P. Warner, ``{Non-BPS Black
  Rings and Black Holes in Taub-NUT},''
  \href{http://dx.doi.org/10.1088/1126-6708/2009/06/015}{{\em JHEP} {\bf 0906}
  (2009)  015}, \href{http://arxiv.org/abs/0902.4526}{{\tt arXiv:0902.4526
  [hep-th]}}.

\bibitem{Bena:2009en}
I.~Bena, S.~Giusto, C.~Ruef, and N.~P. Warner, ``{Multi-Center non-BPS Black
  Holes: the Solution},''
  \href{http://dx.doi.org/10.1088/1126-6708/2009/11/032}{{\em JHEP} {\bf 0911}
  (2009)  032}, \href{http://arxiv.org/abs/0908.2121}{{\tt arXiv:0908.2121
  [hep-th]}}.

\bibitem{Bena:2009fi}
I.~Bena, S.~Giusto, C.~Ruef, and N.~P. Warner, ``{Supergravity Solutions from
  Floating Branes},'' \href{http://dx.doi.org/10.1007/JHEP03(2010)047}{{\em
  JHEP} {\bf 1003} (2010)  047}, \href{http://arxiv.org/abs/0910.1860}{{\tt
  arXiv:0910.1860 [hep-th]}}.

\bibitem{Bobev:2009kn}
N.~Bobev and C.~Ruef, ``{The Nuts and Bolts of Einstein-Maxwell Solutions},''
  \href{http://dx.doi.org/10.1007/JHEP01(2010)124}{{\em JHEP} {\bf 1001} (2010)
   124}, \href{http://arxiv.org/abs/0912.0010}{{\tt arXiv:0912.0010 [hep-th]}}.

\bibitem{Bobev:2011kk}
N.~Bobev, B.~Niehoff, and N.~P. Warner, ``{Hair in the Back of a Throat:
  Non-Supersymmetric Multi- Center Solutions from K\'ahler Manifolds},''
\href{http://arxiv.org/abs/1103.0520}{{\tt arXiv:1103.0520 [hep-th]}}.

\bibitem{Dall'Agata:2010dy}
G.~Dall'Agata, S.~Giusto, and C.~Ruef, ``{U-duality and non-BPS solutions},''
  \href{http://arxiv.org/abs/1012.4803}{{\tt arXiv:1012.4803 [hep-th]}}.

\bibitem{bossard-ruef}
G.~Bossard and C.~Ruef, ``{work in progress},''
  \href{http://arxiv.org/abs/1104.????}{{\tt arXiv:1104.???? [hep-th]}}.

\bibitem{Bates:2003vx}
B.~Bates and F.~Denef, ``{Exact solutions for supersymmetric stationary black
  hole composites},''
\href{http://arxiv.org/abs/hep-th/0304094}{{\tt arXiv:hep-th/0304094}}.

\bibitem{Elvang:2005sa}
H.~Elvang, R.~Emparan, D.~Mateos, and H.~S. Reall, ``{Supersymmetric 4-D
  rotating black holes from 5-D black rings},''
  \href{http://dx.doi.org/10.1088/1126-6708/2005/08/042}{{\em JHEP} {\bf 0508}
  (2005)  042}, \href{http://arxiv.org/abs/hep-th/0504125}{{\tt
  arXiv:hep-th/0504125 [hep-th]}}.

\bibitem{Gaiotto:2005xt}
D.~Gaiotto, A.~Strominger, and X.~Yin, ``{5D black rings and 4D black holes},''
  {\em JHEP} {\bf 02} (2006)  023,
\href{http://arxiv.org/abs/hep-th/0504126}{{\tt arXiv:hep-th/0504126}}.

\bibitem{Bena:2005ni}
I.~Bena, P.~Kraus, and N.~P. Warner, ``Black rings in taub-nut,'' {\em Phys.
  Rev.} {\bf D72} (2005)  084019,
\href{http://arxiv.org/abs/hep-th/0504142}{{\tt hep-th/0504142}}.

\bibitem{Camps:2008hb}
J.~Camps, R.~Emparan, P.~Figueras, S.~Giusto, and A.~Saxena, ``{Black Rings in
  Taub-NUT and D0-D6 interactions},''
  \href{http://dx.doi.org/10.1088/1126-6708/2009/02/021}{{\em JHEP} {\bf 02}
  (2009)  021},
\href{http://arxiv.org/abs/0811.2088}{{\tt arXiv:0811.2088 [hep-th]}}.

\bibitem{Pomeransky:2006bd}
A.~Pomeransky and R.~Sen'kov, ``{Black ring with two angular momenta},''
  \href{http://arxiv.org/abs/hep-th/0612005}{{\tt arXiv:hep-th/0612005
  [hep-th]}}.

\bibitem{Hoskisson:2008qq}
J.~Hoskisson, ``{A Charged Doubly Spinning Black Ring},''
  \href{http://dx.doi.org/10.1103/PhysRevD.79.104022}{{\em Phys.Rev.} {\bf D79}
  (2009)  104022}, \href{http://arxiv.org/abs/0808.3000}{{\tt arXiv:0808.3000
  [hep-th]}}.

\bibitem{Gal'tsov:2009da}
D.~V. Gal'tsov and N.~G. Scherbluk, ``{Three-charge doubly rotating black
  ring},'' \href{http://dx.doi.org/10.1103/PhysRevD.81.044028}{{\em Phys.Rev.}
  {\bf D81} (2010)  044028}, \href{http://arxiv.org/abs/0912.2771}{{\tt
  arXiv:0912.2771 [hep-th]}}.

\bibitem{Bena:2005va}
I.~Bena and N.~P. Warner, ``{Bubbling supertubes and foaming black holes},''
  \href{http://dx.doi.org/10.1103/PhysRevD.74.066001}{{\em Phys.Rev.} {\bf D74}
  (2006)  066001}, \href{http://arxiv.org/abs/hep-th/0505166}{{\tt
  arXiv:hep-th/0505166 [hep-th]}}.

\bibitem{Berglund:2005vb}
P.~Berglund, E.~G. Gimon, and T.~S. Levi, ``{Supergravity microstates for BPS
  black holes and black rings},'' {\em JHEP} {\bf 06} (2006)  007,
\href{http://arxiv.org/abs/hep-th/0505167}{{\tt arXiv:hep-th/0505167}}.

\bibitem{Balasubramanian:2006gi}
V.~Balasubramanian, E.~G. Gimon, and T.~S. Levi, ``{Four Dimensional Black Hole
  Microstates: From D-branes to Spacetime Foam},''
  \href{http://dx.doi.org/10.1088/1126-6708/2008/01/056}{{\em JHEP} {\bf 01}
  (2008)  056},
\href{http://arxiv.org/abs/hep-th/0606118}{{\tt arXiv:hep-th/0606118}}.

\bibitem{Israel:1972vx}
W.~Israel and G.~Wilson, ``{A class of stationary electromagnetic vacuum
  fields},'' \href{http://dx.doi.org/10.1063/1.1666066}{{\em J.Math.Phys.} {\bf
  13} (1972)  865--871}.

\bibitem{Giusto:2004kj}
S.~Giusto and S.~D. Mathur, ``{Geometry of D1-D5-P bound states},''
  \href{http://dx.doi.org/10.1016/j.nuclphysb.2005.09.037}{{\em Nucl. Phys.}
  {\bf B729} (2005)  203--220},
\href{http://arxiv.org/abs/hep-th/0409067}{{\tt arXiv:hep-th/0409067}}.

\end{thebibliography}\endgroup

\end{document}